\documentclass[aps,prd,nofootinbib,amsmath,amssymb,superscriptaddress,twocolumn,10pt]{revtex4}
\usepackage{amsmath}
\usepackage{amsfonts}
\usepackage{txfonts}
\usepackage{graphicx}
\usepackage{dcolumn}
\usepackage{bbm}
\usepackage{amssymb}
\usepackage{latexsym}
\usepackage{CJK}
\usepackage{titlesec}
\usepackage{color}
\usepackage[colorlinks=true, linkcolor=red, citecolor=blue]{hyperref}
\usepackage{appendix}
\let\sb=_ \catcode`\_=\active \def_#1{\ensuremath \sb{\rm#1}}

\begin{document}

 \bibliographystyle{unsrt}

\title{No evidence for the evolution of mass density power-law index $\gamma$ from strong gravitational lensing observation}

\author{Jing-Lei Cui}
\affiliation{Department of Physics, College of Sciences, Northeastern University,
Shenyang 110819, China}
\author{Hai-Li Li}
\affiliation{Department of Physics, College of Sciences, Northeastern University,
Shenyang 110819, China}
\author{Xin Zhang\footnote{Corresponding author}}
\email{zhangxin@mail.neu.edu.cn} \affiliation{Department of Physics, College of Sciences,
Northeastern University, Shenyang 110819, China}
\affiliation{Center for High Energy Physics, Peking University, Beijing 100080, China}

\begin{abstract}
 In this paper, we consider the singular isothermal sphere lensing model that has a spherically symmetric power-law mass distribution $\rho_{tot}(r)\sim r^{-\gamma}$. We investigate whether the mass density power-law index $\gamma$ is cosmologically evolutionary by using the strong gravitational lensing (SGL) observation, in combination with other cosmological observations. We also check whether the constraint result of $\gamma$ is affected by the cosmological model, by considering several simple dynamical dark energy models. We find that the constraint on $\gamma$ is mainly decided by the SGL observation and independent of the cosmological model, and we find no evidence for the evolution of $\gamma$ from the SGL observation.
\end{abstract}


\maketitle

\renewcommand{\thesection}{\arabic{section}}
\renewcommand{\thesubsection}{\arabic{subsection}}
\titleformat*{\section}{\flushleft\bf}
\titleformat*{\subsection}{\flushleft\bf}

\section {Introduction}

Strong gravitational lensing (SGL) observation has been largely developed in recent years and it has the potential to become an important astrophysical tool for probing both cosmology and galaxies. In an SGL system, the ratio of the proper angular diameter distances between lens and source and between observer and source can be measured. Since these distances depend on cosmological geometry, one can thus use the ratio to constrain cosmological models when adequate accurate SGL data can be obtained. For works on the issue concerning the SGL observation, see, e.g., Refs.~\cite{Zhu:2000ee,Chae:2004jp,Zhu:2008sg,Cao:2011bg,Liao:2012ws,Wu:2014wra,Chen:2013vea,Cui:2015oda}.

When calculating the ratio, mass distribution and evolution of the strong-lensing system based on lens model should be considered. Recently, it was shown in Ref.~\cite{Ruff:2010rv} that, when a spherically symmetric power-law total mass-density profile model, $\rho_{tot}(r)\propto r^{-\gamma}$, is considered, a hint of the cosmological evolution of the power-law index $\gamma$ (at around the $2\sigma$ level) can be found. This result is of great interest because the evolution of $\gamma$ is important for the studies of galaxy structure. Thus, more recently, this issue was revisited in Refs.~\cite{Cao:2015qja,Li:2015sla}. In Ref.~\cite{Cao:2015qja}, the singular isothermal sphere (SIS) lens model is adopted and the power-law index $\gamma$ is taken as a free parameter fitted together with cosmological parameters by using a reasonable sample of $118$ strong lenses. But, in this work, when estimating parameters, the present-day fractional energy density of matter $\Omega_{m0}$ is fixed at $0.315$. Furthermore, in Ref.~\cite{Li:2015sla}, in order to precisely constrain $\Omega_{m0}$, the SGL observation is combined with the baryonic acoustic oscillations (BAO) and type Ia supernovae (SNIa) observations. In this study~\cite{Li:2015sla}, the result of $\gamma_0=2.094^{+0.053}_{-0.056}$ and $\gamma_1=-0.053^{+0.103}_{-0.102}$ is obtained based on the $\Lambda$CDM model, for the parametrization $\gamma(z)=\gamma_0+\gamma_1z$, with the combination of SGL and BAO data, which implies that a time-varying power-law index $\gamma$ is not supported.

The aim of the present paper is to carefully check whether the evidence of the cosmological evolution of $\gamma$ can be found from the analysis of SGL observation (in combination with other various cosmological observations). We take two key steps to make the analysis: (i) we constrain the time-varying $\gamma$ (parametrized in the form of $\gamma(z)=\gamma_0+\gamma_1z$) in the $\Lambda$CDM model by using the SGL data combined with different cosmological observations, and (ii) we further check if the constraint result is relevant to the cosmological model considered in the analysis. The current cosmological observations used in this work include: the SNIa data from the ``joint light-curve analysis" (JLA) compilation, the cosmic microwave background (CMB) anisotropy data from the Planck 2015 observation, the BAO data from the 6dFGS, SDSS-DR7, and BOSS-DR11 surveys, and the Hubble parameter measurements (see Ref.~\cite{Guo:2015gpa} for a summary of the $H(z)$ data). The typical dark energy models considered in this work include: the constant $w$ dark energy model (also abbreviated as the $w$CDM model), the holographic dark energy (HDE) model~\cite{Li:2004rb}, and the Ricci dark energy (RDE) model~\cite{Gao:2007ep}, all of which are the models with one more parameter than the $\Lambda$CDM model.

This paper is organized as follows. In Section~\ref{sec.2}, we briefly describe the observational data. In Section~\ref{sec.3}, we constrain the time-varying $\gamma$ in the $\Lambda$CDM model by using the different combinations of SGL with other observations. In Section~\ref{sec.4}, we test the constraint results in dynamical dark energy models. Conclusion is given in Section~\ref{sec.5}.

\section {Observational data}\label{sec.2}

In this section, we briefly describe the observational data used in this work, including SGL, SNIa, CMB, BAO, and $H(z)$ measurements. Throughout this paper, we consider a spatially flat Friedmann-Robertson-Walker (FRW) universe consisting of dark energy (de), matter (m) and radiation (r).

\subsection*{2.1 Strong gravitational lensing}

In strong lensing system, light rays from a source to us can be bent when they pass a massive galaxy or galaxy cluster acting as lens, which forms multiple images of the source, arcs or Einstein rings. Mass distribution model within the lens is related to the parameters of lensing system directly or indirectly. In this paper, we consider the SIS model generalized to the spherically symmetric power-law mass distribution $\rho_{tot}(r)\sim r^{-\gamma}$, following Ref.~\cite{Cao:2015qja}. To make SGL probe usable in parameter estimation, the Einstein radius $\theta_E$ in an SIS lens must be used, which is expressed as
\begin{equation}
\theta_E = 4\pi \frac{\sigma^2_{\rm{ap}}}{c^2}\frac{D_{\rm{A}}(z_l,z_s)}{D_{\rm{A}}(0,z_s)}\left(\frac{\theta_E}{\theta_{ap}}\right)^{2-\gamma}f(\gamma),\label{theta}
\end{equation}
where
\begin{equation}
\begin{aligned}
f(\gamma)=\frac{(5-2\gamma)(\gamma-1)}{\sqrt{\pi}(3-\gamma)}\frac{\Gamma(\gamma-1)}{\Gamma(\gamma-3/2)}\left[\frac{\Gamma(\gamma/2-1/2)}{\Gamma(\gamma/2)}\right]^2\\
\text{($\Gamma(x)$ is gamma function)},
\end{aligned}
\end{equation}
 $\sigma_{\rm{ap}}$ is the velocity dispersion inside an aperture with size $\theta_{ap}$, $\gamma$ is the mass density power-law index, $D_{\rm{A}}(z_l,z_s)$ and $D_{\rm{A}}(0,z_s)$ are the proper angular diameter distances between lens and source and between observer and source, respectively. The proper angular diameter distance in a flat FRW universe is given by
\begin{equation}
D_{\rm{A}}(0,z)=\frac{1}{H_0(1+z)}\int_0^z \frac{dz'}{E(z')},\label{DA}
\end{equation}
where $E(z)\equiv H/H_0$ is the dimensionless Hubble expansion rate, depending on the specific cosmological model, and $H_{0}=100h$ km s$^{-1}$ Mpc$^{-1}$ is the Hubble constant ($h$ is the reduced Hubble constant). To constrain cosmological models, we use the distance ratio
\begin{equation}
\mathcal{D}(z_{\rm{l}},z_{\rm{s}}) = \frac{D_{\rm{A}}(z_{\rm{l}},z_{\rm{s}})}{D_{\rm{A}}(0,z_{\rm{s}})}=\frac{\int_{z_{\rm{l}}}^{z_{\rm{s}}} dz'/E(z')}{\int_0^{z_{\rm{s}}} dz'/E(z')}.
\end{equation}
The observable $\mathcal{D}^{\rm{obs}}$ can be gained from Eq.~(\ref{theta}). Accordingly, the uncertainty of $\mathcal{D}^{obs}$ is
\begin{equation}
\sigma_{\mathcal{D}}=\mathcal{D}^{\rm{obs}}\sqrt{4\left(\frac{\sigma_{\sigma_{ap}}}{\sigma_{ap}}\right)^2+(1-\gamma)^2\left(\frac{\sigma_{\theta_E}}{\theta_E}\right)^2}.
\end{equation}
Here, it is assumed that the fractional uncertainty of the Einstein radius is at the level of $5\%$, i.e., $\frac{\sigma_{\theta_E}}{\theta_E}=0.05$ for all the lenses.

In the cosmological fit, we replace $\sigma_{ap}$ with the aperture-corrected velocity dispersion $\sigma_0$ ($\sigma_0=\sigma_{ap}(\theta_{eff}/(2\theta_{ap}))^{-0.04}$, where $\theta_{eff}$ is the effective radius). We use $\sigma_0$ to make the observable $\mathcal{D}^{\rm{obs}}$ more homogeneous for the sample of lenses located at different redshifts, and to make the fitting results more consistent with the previous analysis; see Ref.~\cite{Cao:2015qja} for more details. Moreover, we consider mass density power-law index $\gamma$ evolving with redshift, i.e., $\gamma=\gamma_0+\gamma_1z$, where $\gamma_0$ and $\gamma_1$ are free parameters. We use data sample of $118$ strong lenses from Table $1$ of Ref.~\cite{Cao:2015qja} compiled from the Sloan Lens ACS Survey (SLACS)~\cite{Bolton:2008xf,Auger:2009hj}, BOSS emission-line lens survey (BELLS)~\cite{Bolton:2011bj}, Lens Structure and Dynamics (LSD)~\cite{Treu:2002ee,Koopmans:2002qh,Treu:2004wt}, and Strong Lensing Legacy Survey (SL2S)~\cite{Sonnenfeld:2013cha,Sonnenfeld:2013xga}.

We fit the parameters by minimizing the following $\chi^2$ function,
\begin{equation}
\chi^2_{\rm{SGL}}=\sum\limits_{i}\frac{(\mathcal{D}^{\rm{th}}_i-\mathcal{D}^{\rm{obs}}_i)^2}{\sigma^2_{\mathcal{D},i}}.
\end{equation}

\subsection*{2.2 Type Ia supernova}
We use the JLA compilation of $740$ spectroscopically confirmed SNIa with high quality light curves~\cite{Betoule:2014frx}. Theoretically, the relation of standardized distance modulus $\mu$ to luminosity distance $d_L$ for an SNIa is given by
\begin{equation}
\mu_{th}=5\log_{10}\frac{d_L}{10\rm{pc}},
\end{equation}
and
\begin{equation}
d_L(z,z_{hel})=\frac{1+z_{hel}}{H_0} \int_{0}^{z} \frac{dz'}{E(z')},
\end{equation}
where $z$ and $z_{hel}$ denote the CMB frame and heliocentric redshifts, respectively.
 In the JLA analysis, the distance modulus of an SNIa is quantified by an empirical linear model,
\begin{equation}
\mu=m^{\star}_B-(M_{B}-\alpha \times X_1+\beta \times C),
\end{equation}
where $m^{\star}_B$ is the observed peak magnitude in the rest frame B band, $M_B$ is the absolute magnitude, $X_1$ discribes the time stretching of the light curve, $C$ is the supernova color at maximum brightness. The two nuisance parameters $\alpha$ and $\beta$ are assumed to be constants for all SNIa. For the JLA supernova, the $\chi^2$ function is
\begin{equation}
\chi^{2}_{SN}=(\mu-\mu_{th})^{\dagger}\rm{C}^{-1}_{SN}(\mu-\mu_{th}),
\end{equation}
where $\rm{C}_{SN}$ is the covariance matrix of $\mu$, and can be found in Ref.~\cite{Betoule:2014frx}.

\subsection*{2.3 Cosmic microwave background}

We adopt the ``Planck distance priors" from the Planck 2015 observation~\cite{Ade:2015rim}. The distance priors include the shift parameter $R$, the acoustic scale $\ell_{\it a}$ and the baryon density $\omega_b$. They are respectively defined as
\begin{equation}
R\equiv \sqrt{\Omega_{m0}H^2_0}(1+z_*)D_A(z_*),
\end{equation}
\begin{equation}
\ell_{\it a} \equiv (1+z_*)\frac{\pi D_A(z_*)}{r_s(z_*)},
\end{equation}
and
\begin{equation}
\omega_b\equiv\Omega_{b0}h^2,
\end{equation}
where $\Omega_{b0}$ is the present-day fractional energy density of baryon, $D_{\rm{A}}(z_*)$ is the proper angular diameter distance at the redshift of the decoupling epoch of photon $z_*$, and $r_{\rm{s}}(z_*)$ is the comoving sound horizon at the photon-decoupling epoch. Here, $r_{\rm{s}}(z)$ is given by
\begin{equation}
r_s(z)=H^{-1}_0\int^a_0 \frac{da'}{a'^2E(a')\sqrt{3(1+\overline{R_b}a')}},\label{rs}
\end{equation}
where $\overline{R_{\rm{b}}}a=3\rho_{\rm{b0}}/(4\rho_{\rm{\gamma0}})$ with $\rho_{\rm{b0}}$ and $\rho_{\rm{\gamma0}}$ being the present-day baryon and photon energy densities, respectively, $\overline{R_{\rm{b}}}=31500\Omega_{\rm{b0}}h^2(T_{\rm{cmb}}/2.7\rm{K})^{-4}$, and $T_{\rm{cmb}}=2.7255\rm{K}$. $z_*$ is given by the fitting formula~\cite{Hu:1995en},
\begin{equation}
z_*=1048[1+0.00124(\Omega_{\rm{b0}}h^2)^{-0.738}][1+g_1(\Omega_{\rm{m0}}h^2)^{g_2}],
\end{equation}
where
\begin{equation}
 g_1=\frac{0.0783(\Omega_{\rm{b0}}h^2)^{-0.238}}{1+39.5(\Omega_{\rm{b0}}h^2)^{-0.763}},~~g_2=\frac{0.560}{1+21.1(\Omega_{\rm{b0}}h^2)^{1.81}}.
\end{equation}

We use the mean values and covariance matrix of $\{\ell_{\it{a}},R,\omega_b\}$ for the Planck TT+LowP data from Ref.~\cite{Ade:2015rim}. The $\chi^2$ function for CMB is
\begin{equation}
\chi^2_{CMB}=\Delta p_i[{\rm{Cov}}^{-1}_{CMB}(p_i,p_j)]\Delta p_j,~~~\Delta p_i=p^{\rm{th}}_i-p^{\rm{obs}}_i,
\end{equation}
where $p_1=\ell_{\it{a}}$, $p_2=R$, and $p_3=\omega_{\rm{b}}$.

\subsection*{2.4 Baryon acoustic oscillations}

Through BAO measurements, one can get the ratio of the effective distance measure $D_{\rm{V}}(z)$ and the comoving sound horizon $r_{\rm{s}}(z_d)$. The spherical average gives us the expression of $D_{\rm{V}}(z)$,
\begin{equation}
D_{\rm{V}}(z)\equiv \left [(1+z)^2D^2_{\rm{A}}(z)\frac{z}{H(z)}\right ]^{1/3}.
\end{equation}
The comoving sound horizon size $r_{\rm{s}}(z_d)$ can be calculated by using Eq.~(\ref{rs}), where $z_d$ denotes the redshift of the drag epoch, whose fitting formula is given by~\cite{Eisenstein:1997ik},
\begin{equation}
z_d=\frac{1291(\Omega_{\rm{m0}}h^2)^{0.251}}{1+0.659(\Omega_{\rm{m0}}h^2)^{0.828}}[1+b_1(\Omega_{\rm{b0}}h^2)^{b_2}],
\end{equation}
where
\begin{equation}
\begin{gathered}
b_1=0.313(\Omega_{\rm{m0}}h^2)^{-0.419}[1+0.607(\Omega_{\rm{m0}}h^2)^{0.674}],\\
b_2=0.238(\Omega_{\rm{m0}}h^2)^{0.223}.
\end{gathered}
\end{equation}

We use four BAO data points from the 6dF Galaxy Survey~\cite{Beutler:2011hx}, the SDSS-DR7~\cite{Ross:2014qpa} and the BOSS-DR11~\cite{Anderson:2013zyy} surveys. The $\chi^{2}$ function for BAO is
\begin{equation}
\chi^{2}_{\rm{BAO}}=\Delta p_{i}[{\rm Cov}^{-1}_{\rm{BAO}}(p_{i},p_{j})]\Delta p_{j}, \; \Delta p_{i}=p_{i}^{\rm{th}}-p_{i}^{\rm{obs}}.
\end{equation}
The concrete information of $\Delta p_{i}$ corresponding to the four BAO data points is explicitly given in Ref.~\cite{Xu:2016grp} (see also Refs.~\cite{Cheng:2014kja,Hu:2015bpa}).

\subsection*{2.5 Hubble parameter}

Measuring directly the Hubble parameter $H(z)$ is very important for exploring the history of cosmic evolution. Reference~\cite{Guo:2015gpa} sorted out $31$ $H(z)$ data from the measurements of clustering of galaxies or quasars~\cite{Anderson:2013zyy,Blake:2012pj,Chuang:2012qt,Delubac:2014aqe} and differential age~\cite{Stern:2009ep,Moresco:2012jh,Zhang:2012mp,Moresco:2015cya}. In Table 1 of Ref.~\cite{Guo:2015gpa}, the data of $H(z)$ versus the redshift $z$ and corresponding errors are listed. The $\chi^{2}$ function for $H(z)$ is
\begin{equation}
\chi^{2}_{\rm{H}}=\sum\limits_{i=1}^{N}\frac{[H^{\rm{th}}-H^{\rm{obs}}]^2}{\sigma_i^2},
\end{equation}
where $H^{\rm{th}}=H_0E(z)$.

\section {Constraining time-varying $\gamma$ in the $\Lambda$CDM model}\label{sec.3}

\begin{table}\tiny
\caption{Constraint results in the $\Lambda$CDM model by using the SGL data, in combination with JLA, CMB, BAO, and $H(z)$. For convenience, we use JCBH as an abbreviation to denote JLA+CMB+BAO+$H(z)$. }
\label{table1}
\small
\setlength\tabcolsep{2.8pt}
\renewcommand{\arraystretch}{1.5}
\centering
\begin{tabular}{ccccccccc}
\\
\hline\hline

  Observation & $\gamma_0$ & $\gamma_1$ & $\Omega_{m0}$ &$h$  \\ \hline
SGL
                   & $2.142^{+0.037}_{-0.057}$
                   & $-0.053^{+0.089}_{-0.093}$
                   & $0.187^{+0.133}_{-0.114}$
                   & $0.340^{+0.660}_{-0.040}$ \\
SGL+JLA
                   & $2.103^{+0.045}_{-0.054}$
                   & $-0.062^{+0.096}_{-0.105}$
                   & $0.328^{+0.025}_{-0.029}$
                   & $0.572^{+0.428}_{-0.271}$\\
SGL+CMB
                   & $2.111^{+0.040}_{-0.056}$
                   & $-0.071^{+0.104}_{-0.097}$
                   & $0.313^{+0.013}_{-0.014}$
                   & $0.675^{+0.010}_{-0.009}$\\
SGL+BAO
                   & $2.109^{+0.046}_{-0.054}$
                   & $-0.065^{+0.097}_{-0.102}$
                   & $0.305^{+0.063}_{-0.058}$
                   & $0.591^{+0.409}_{-0.142}$\\
SGL+$H(z)$
                   & $2.123^{+0.039}_{-0.049}$
                   & $-0.061^{+0.090}_{-0.099}$
                   & $0.262^{+0.030}_{-0.028}$
                   & $0.707^{+0.024}_{-0.023}$\\
SGL+JCBH
                   & $2.101^{+0.047}_{-0.045}$
                   & $-0.058^{+0.091}_{-0.106}$
                   & $0.319^{+0.008}_{-0.006}$
                   & $0.671^{+0.005}_{-0.006}$\\   \hline\hline

\end{tabular}
\end{table}

In this section, we constrain the time-varying $\gamma$, in the form of $\gamma(z)=\gamma_0+\gamma_1z$, along with other cosmological parameters, in the $\Lambda$CDM model, by using the different combinations of SGL with other cosmological observations.

For the $\Lambda$CDM model, the dimensionless Hubble expansion rate $E(z)$ is expressed as
\begin{equation}
E(z)=\left[\Omega_{m0}(1+z)^{3}+\Omega_{r0}(1+z)^{4}+(1-\Omega_{m0}-\Omega_{r0})\right]^{1/2},
\end{equation}
where we have the fractional density of radiation $\Omega_{\rm{r0}}=2.469\times10^{-5}h^{-2}(1+0.2271N_{\rm{eff}})$ and the effective number of neutrino species $N_{\rm{eff}}=3.046$.

In this analysis, we consider the cases of SGL alone, SGL+JLA, SGL+CMB, SGL+BAO, SGL+$H(z)$, and SGL+JCBH. Here, for convenience, we use the abbreviation JCBH to denote the combination of JLA+CMB+BAO+$H(z)$. The constraint results of $\gamma_0$, $\gamma_1$, $\Omega_{m0}$, and $h$ are summarized in Table~\ref{table1}. From the table, we find that the constraint results of all the data combinations are consistent with each other (except for that the SGL alone cannot precisely constrain $\Omega_{m0}$ and $h$). In all the cases, we find that $\gamma_0\approx 2.1$ (with the uncertainty around 0.04--0.05) and $\gamma_1\approx -0.06$ (with the uncertainty around 0.1), which indicates that the time-varying $\gamma$ is not favored by the current observations.

\begin{figure}[htbp]
\centering
\includegraphics[scale=0.35]{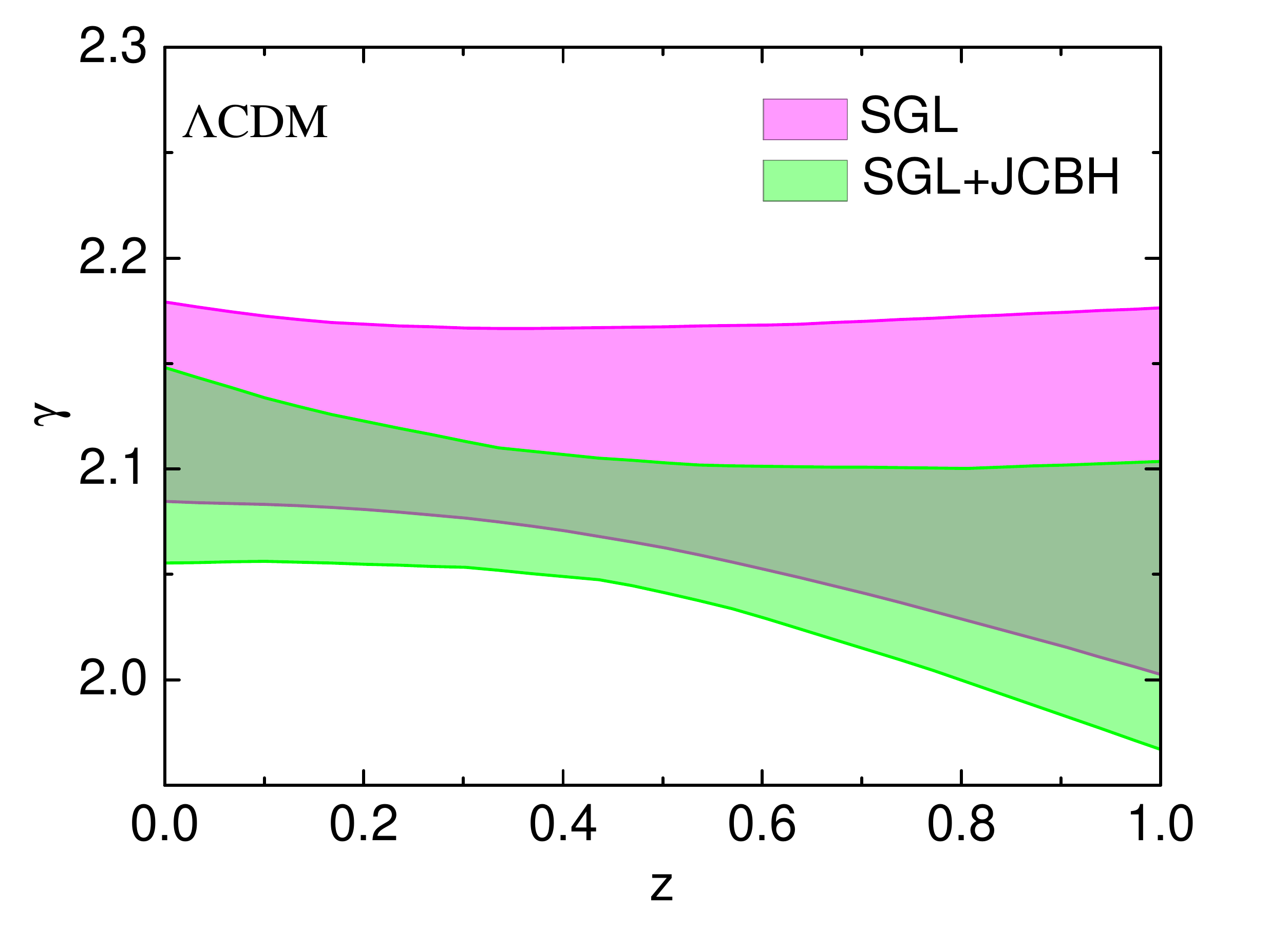}
\caption{\label{fig1} Reconstruction of $\gamma(z)$ (with 1$\sigma$ error) in the $\Lambda$CDM model from SGL alone and SGL+JCBH.}
\end{figure}

We find that the SGL data alone can constrain $\gamma$ tightly, and when other cosmological observations are considered, the constraint results are not affected greatly. This indicates that $\gamma$ is indeed an astrophysical parameter, without cosmological evolution. For all the cases, the result is consistent with $\gamma_1=0$ at round the 0.6$\sigma$ level. In Fig.~\ref{fig1}, we reconstruct the $\gamma(z)$ evolution according to the constraint results of the cases SGL alone and SGL+JCBH. From the figure, we clearly see that the two cases are well consistent with each other, both in good agreement with a constant $\gamma$ (with $\gamma_1=0$).

From Table~\ref{table1}, we can clearly see that using the SGL data alone cannot strictly constrain $\Omega_{m0}$ and $h$. When the SGL observation is combined with JCBH, the results of $\Omega_{m0}$ and $h$ are consistent with those from Planck $2015$ ($\Omega_m=0.308\pm0.012$ and $h=0.678\pm0.009$)~\cite{Ade:2015xua}. In the case of SGL+JCBH, we obtain $\gamma_0=2.101^{+0.047}_{-0.045}$ and $\gamma_1=-0.058^{+0.091}_{-0.106}$, which is the fit result of $\gamma$ from the current cosmological observations.

\section {Constraints in dynamical dark energy models}\label{sec.4}

\begin{table*}\tiny
\caption{Constraint results for the $w$CDM, HDE and RDE models by using SGL alone and SGL+JCBH.}
\label{table2}
\small
\setlength\tabcolsep{2.8pt}
\renewcommand{\arraystretch}{1.5}
\centering
\begin{tabular}{ccccccccccccccc}
\\
\hline\hline Model&\multicolumn{2}{c}{$w$CDM} &&\multicolumn{2}{c}{HDE}&&\multicolumn{2}{c}{RDE} \\
          \cline{1-1}\cline{3-4}\cline{6-7}\cline{9-10}
Observation   && SGL & SGL+JCBH && SGL&  SGL+JCBH &&  SGL &  SGL+JCBH \\ \hline
$\gamma_0$
                   && $2.151^{+0.043}_{-0.061}$
                   & $2.108^{+0.035}_{-0.057}$ &
                   & $2.149^{+0.046}_{-0.058}$
                   & $2.113^{+0.033}_{-0.066}$ &
                   & $2.152^{+0.042}_{-0.061}$
                   & $2.109^{+0.043}_{-0.067}$ \\
$\gamma_1$
                   && $-0.066^{+0.094}_{-0.111}$
                   & $-0.075^{+0.106}_{-0.085}$ &
                   & $-0.069^{+0.098}_{-0.112}$
                   & $-0.078^{+0.103}_{-0.090}$&
                   & $-0.067^{+0.095}_{-0.110}$
                   & $-0.099^{+0.105}_{-0.096}$\\
$\Omega_m$
                   && $0.207^{+0.145}_{-0.124}$
                   & $0.324^{+0.009}_{-0.007}$&
                   & $0.189^{+0.136}_{-0.121}$
                   & $0.323^{+0.008}_{-0.008}$&
                   & $0.160^{+0.126}_{-0.097}$
                   & $0.344^{+0.006}_{-0.007}$\\
$h$
                   && $0.608^{+0.392}_{-0.308}$
                   & $0.664^{+0.006}_{-0.009}$&
                   & $0.485^{+0.515}_{-0.185}$
                   & $0.657^{+0.007}_{-0.006}$&
                   & $0.551^{+0.449}_{-0.251}$
                   & $0.666^{+0.005}_{-0.004}$\\ \hline
Model parameter
                   && $w=-1.186^{+0.417}_{-0.676}$
                   & $w=-0.968^{+0.035}_{-0.028}$&
                   & $c=0.702^{+0.796}_{-0.404}$
                   & $c=0.741^{+0.036}_{-0.040}$&
                   & $\alpha=0.441^{+0.164}_{-0.137}$
                   & $\alpha=0.333^{+0.009}_{-0.009}$\\  \hline\hline
\end{tabular}
\end{table*}

In this section, we check if the constraint result of $\gamma$ derived in the last section is affected by the cosmological model. Thus, we consider several dynamical dark energy models to make a test. We only consider the simplest models, i.e., the $w$CDM model, the HDE model, and the RDE model, all of which have only one more parameter than $\Lambda$CDM.

For the $w$CDM model, we have $w=$ constant, and thus $E(z)$ is written as
\begin{equation}
E(z)=
\sqrt{\Omega_{\rm{m0}}(1+z)^3+\Omega_{\rm{r0}}(1+z)^4+(1-\Omega_{\rm{m0}}-\Omega_{\rm{r0}})(1+z)^{3(1+w)}}.
\end{equation}

For the HDE model, the density of dark energy is defined by \cite{Li:2004rb}
\begin{equation}
\rho _{de} = 3c^2M^2_{\rm{P}}L^{-2},\label{eq1}
\end {equation}
where $c$ is a dimensionless parameter, $M_{\rm{P}}$ is the reduced Planck mass defined by $M^2_{\rm{P}} = (8\pi G)^{-1}$, and $L$ is the infrared (IR) cutoff, given by the future event horizon of the universe,
\begin{equation}
L=a\int_{\it t}^\infty\frac{dt'}{a}=a\int_{\it a}^\infty\frac{da'}{Ha'^2},\label{eq2}
\end {equation}
with $a$ the scale factor of the universe. The cosmological evolution of this model is governed by the following differential equations:
\begin{equation}
\frac{1}{E}\frac{dE}{dz}=-\frac{\Omega_{\rm{de}}}{1+z}\left(\frac{1}{2}+\frac{\sqrt{\Omega_{\rm{de}}}}{c}-\frac{\Omega_{\rm{r}}+3}{2\Omega_{\rm{de}}}\right)\label{Ez}
\end{equation}
and
\begin{equation}
\frac{d\Omega_{\rm{de}}}{dz}=-\frac{2\Omega_{\rm{de}}(1-\Omega_{\rm{de}})}{1+z}\left(\frac{1}{2}+\frac{\sqrt{\Omega_{\rm{de}}}}{c}+\frac{\Omega_{\rm{r}}}{2(1-\Omega_{\rm{de}})}\right),\label{Ode}
\end{equation}
where $\Omega_{\rm{de}}$ is the fractional density of dark energy and $\Omega_{\rm{r}}=\Omega_{\rm{r0}}(1+z)^4/E(z)^2$. For extensive studies on the HDE model, see, e.g., Refs.~\cite{Zhang:2005yz,Zhang:2005hs,Zhang:2007sh,Zhang:2006av,Zhang:2006qu,Zhang:2009xj,Wang:2013zca,Zhou:2016rtz,Zhang:2015uhk,Wang:2016tsz,Zhang:2017rbg,He:2016rvp}.

\begin{figure}
  \centering
      \label{fig:subfig:a}
    \includegraphics[width=3.5in]{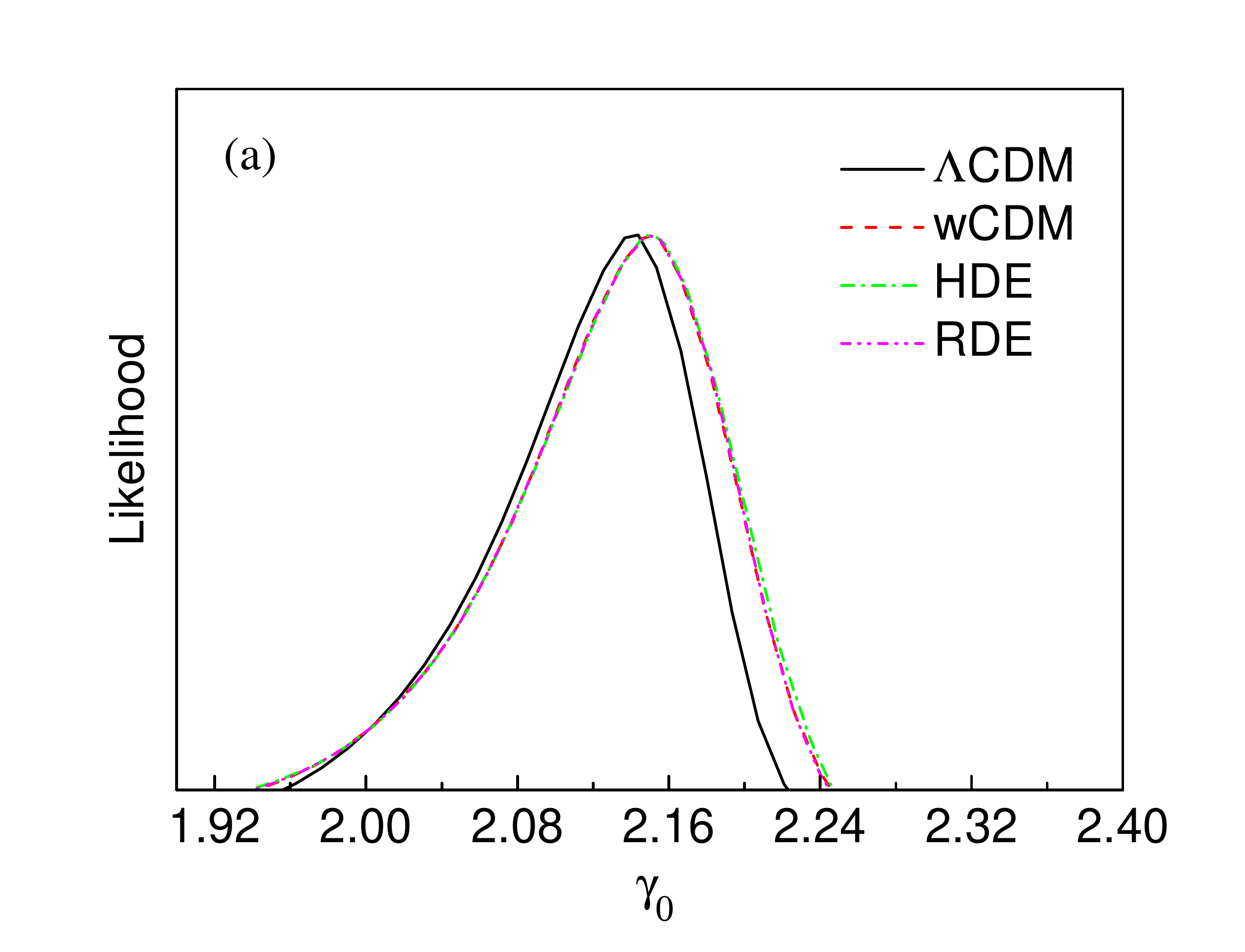}
  \hspace{1in}

    \label{fig:subfig:b}
    \includegraphics[width=3.5in]{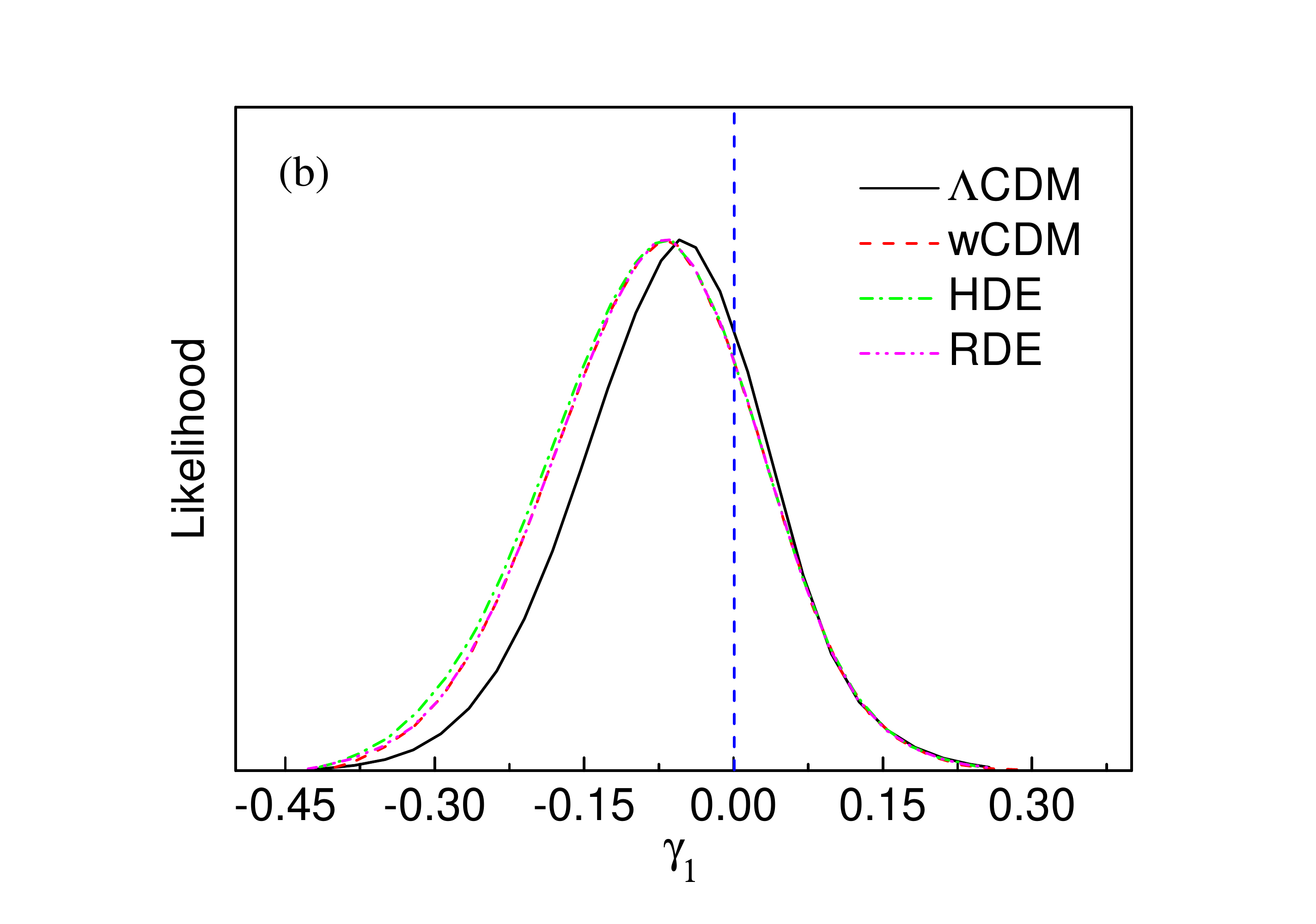}
  \caption{The 1D marginalized probability distributions of $\gamma_0$ and $\gamma_1$ in the $\Lambda$CDM, $w$CDM, HDE and RDE models by using the SGL data alone.}
  \label{fig2} 
\end{figure}

For the RDE model, the density of dark energy is also given by Eq. (\ref{eq1}). The different point is that the IR cutoff $L$ is related to the Ricci scalar curvature, $\mathcal{R}=-6(\stackrel{\centerdot}{H}+2H^2)$~\cite{Gao:2007ep,Zhang:2009un,Feng:2009hr,Li:2009bn,Zhang:2010im,Fu:2011ab,Zhang:2014sqa}. Accordingly, in this model, dark energy density is defined as \cite{Gao:2007ep}
\begin{equation}
\rho_{de}=3\alpha M^2_p(\stackrel{\centerdot}{H}+2H^2),
\end{equation}
where $\alpha$ is a positive constant. Then we can get
\begin{equation}
E(z)=\sqrt{\frac{2\Omega_{\rm{m0}}}{2-\alpha}(1+z)^3+\Omega_{\rm{r0}}(1+z)^4+(1-\frac{2\Omega_{\rm{m0}}}{2-\alpha}-\Omega_{\rm{r0}})(1+z)^{4-\frac{2}{\alpha}}}.
\end{equation}

The constraint results for the $w$CDM, HDE, and RDE models are listed in Table~\ref{table2}. From Table~\ref{table2}, we find that the best-fitting $\gamma_0$ and $\gamma_1$ as well as their corresponding errors for the three models are almost the same to each other when the SGL data only is used. Furthermore, by comparing Table~\ref{table2} with Table~\ref{table1}, we can see that the results of $\gamma_0$ and $\gamma_1$ are also very similar to those of the $\Lambda$CDM model. We plot the 1D marginalized probability distributions of $\gamma_0$ and $\gamma_1$ for the $\Lambda$CDM, $w$CDM, HDE, and RDE models in Fig.~\ref{fig2}. We can clearly see that the curves of the dynamical dark energy models are almost degenerate and their deviation from the result of the $\Lambda$CDM model is also very small.

\begin{figure}
  \centering
      \label{fig:subfig:a}
    \includegraphics[width=3.5in]{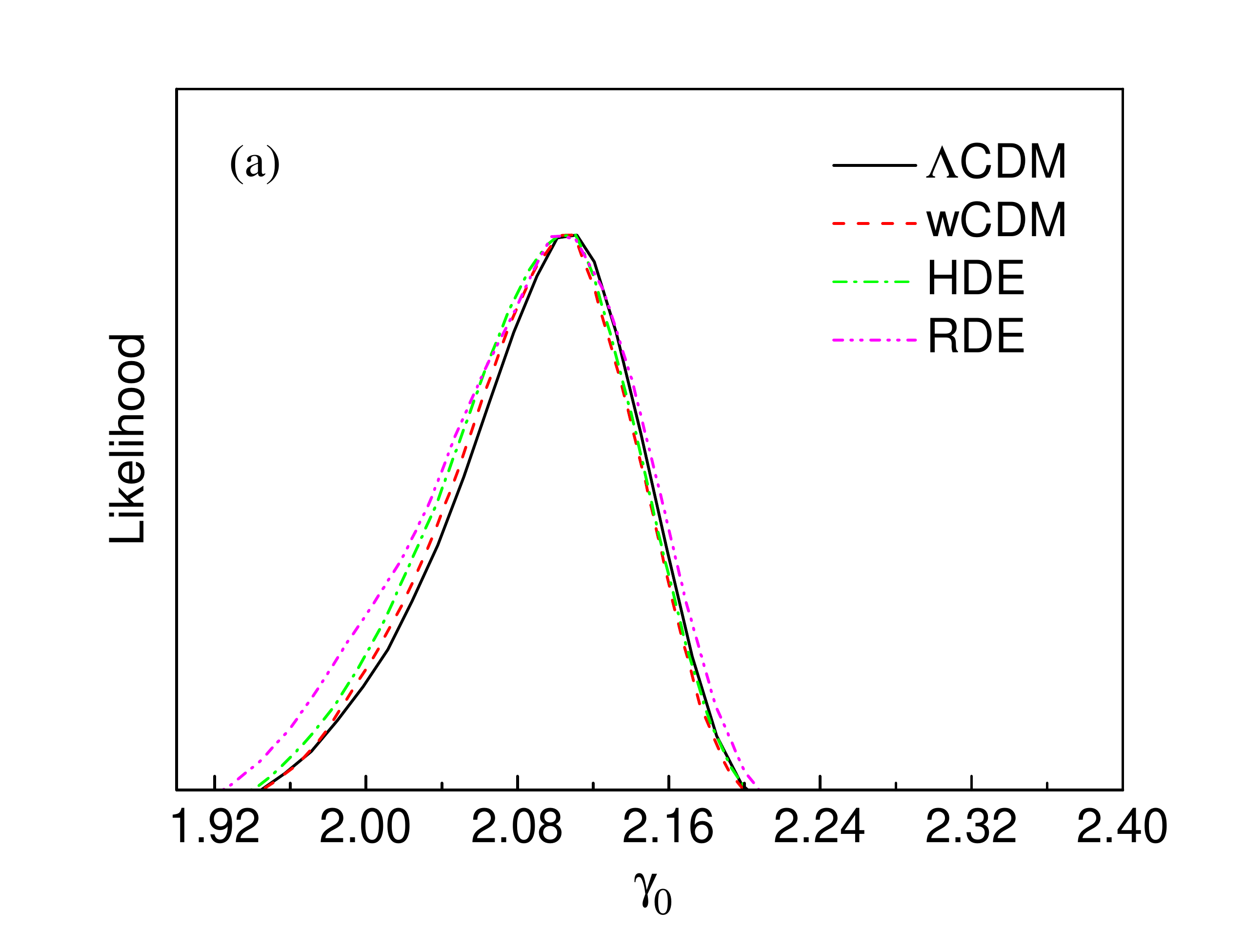}
  \hspace{1in}

    \label{fig:subfig:b}
    \includegraphics[width=3.5in]{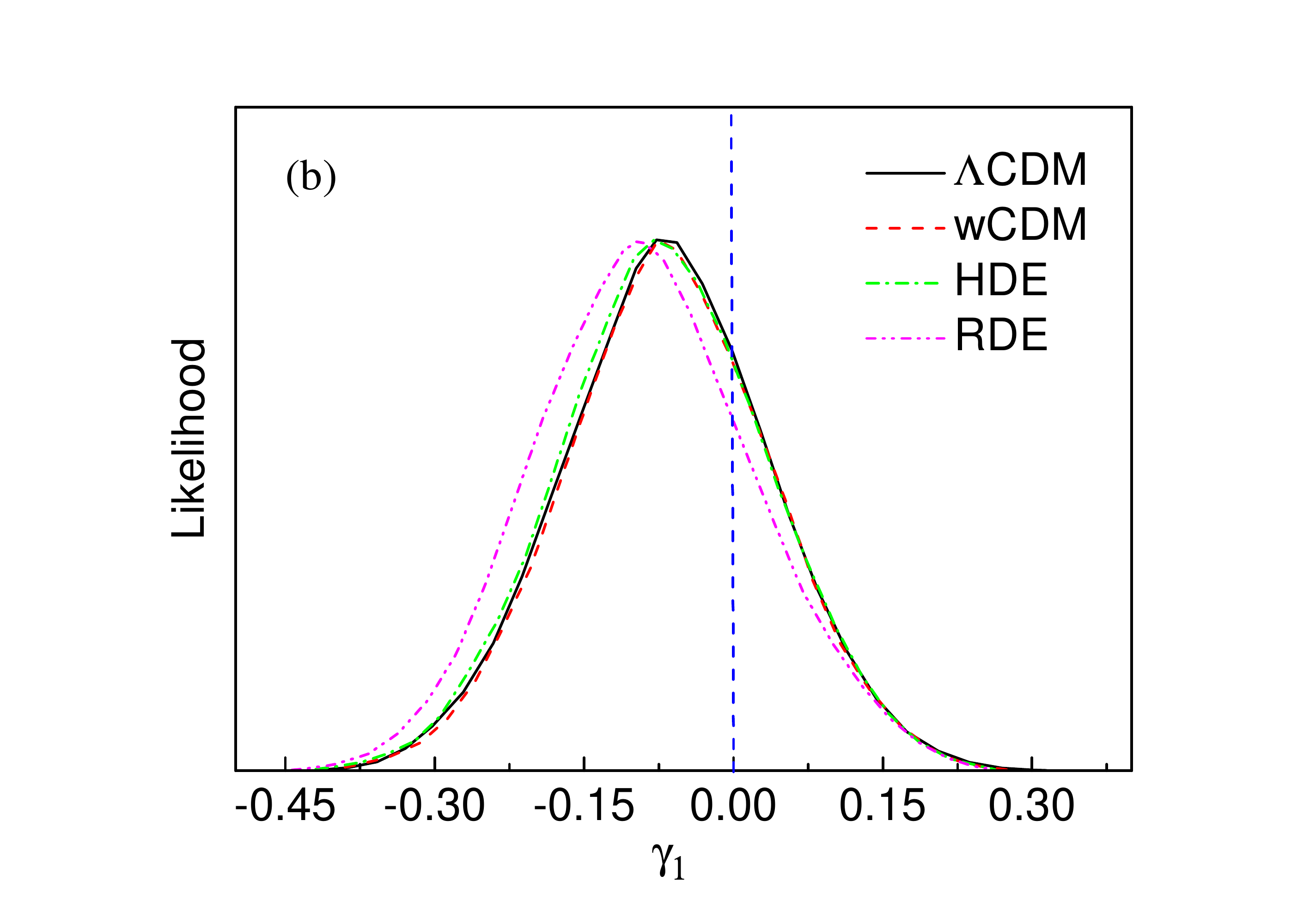}
  \caption{The 1D marginalized probability distributions of $\gamma_0$ and $\gamma_1$ in the $\Lambda$CDM, $w$CDM, HDE and RDE models by using the SGL+JCBH data.}
  \label{fig3} 
\end{figure}

When the combination of SGL and JCBH is used in parameter estimation, we find that the changes for the fit results of $\gamma_0$ and $\gamma_1$, compared to the case of SGL alone, are very small. In this case, the 1D marginalized probability distributions of $\gamma_0$ and $\gamma_1$ for the $\Lambda$CDM, $w$CDM, HDE and RDE models are plotted in Fig.~\ref{fig3}. We also find that the curves are very close to each other. From Table~\ref{table2}, we find that using SGL data alone cannot tightly constrain the cosmological parameters, such as $\Omega_m$ and $h$, and model parameters ($w$, $c$, and $\alpha$). To constrain these parameters well, the SGL data must be combined with other cosmological observations. Our analysis shows that the cosmological model doest not affect the constraint result of $\gamma$.

\section {Conclusion}\label{sec.5}

In this paper, we wish to clarify whether the mass density power-law index $\gamma$ is cosmologically evolutionary by using the SGL observation, in combination with other cosmological observations. We constrain the time-varying $\gamma$, in the form of $\gamma(z)=\gamma_0+\gamma_1z$, in the $\Lambda$CDM model, with the data combinations SGL alone, SGL+JLA, SGL+CMB, SGL+BAO, SGL+$H(z)$, and SGL+JCBH. We find that the constraint results of all the data combinations are consistent with each other, and in all the cases we derive $\gamma_0\approx 2.1$ (with the uncertainty around 0.04--0.05) and $\gamma_1\approx -0.06$ (with the uncertainty around 0.1), which indicates that the time-varying $\gamma$ is not supported by the current observations. Our result is consistent with $\gamma_1=0$ at round the 0.6$\sigma$ level. We then check whether the constraint result of $\gamma$ is affected by the cosmological model, by considering the $w$CDM, HDE, and RDE models. We find that the constraint result of $\gamma$ is not relevant to the cosmological model. Therefore, we conclude that there is no evidence for the cosmological evolution of $\gamma$ from the current SGL observation.

\section*{Acknowledgements}

This work was supported by the National Natural Science Foundation of China (Grants No.~11522540 and No.~11690021), the Top-Notch Young Talents Program of China, and the Provincial Department of Education of Liaoning (Grant No.~ L2012087).


\begin{thebibliography}{}

\bibitem{Zhu:2000ee}
  Z.~H.~Zhu,  
  Mod.\ Phys.\ Lett.\ A {\bf 15} (2000) 1023  [astro-ph/0010351].

\bibitem{Chae:2004jp}
  K.~H.~Chae, G.~Chen, B.~Ratra and D.~W.~Lee,  
  Astrophys.\ J.\  {\bf 607} (2004) L71  [astro-ph/0403256].

\bibitem{Zhu:2008sg}
  Z.~H.~Zhu and M.~Sereno,  
  Astron.\ Astrophys.\  {\bf 487} (2008) 831  [arXiv:0804.2917 [astro-ph]].

\bibitem{Cao:2011bg}
  S.~Cao, Y.~Pan, M.~Biesiada, W.~Godlowski and Z.~H.~Zhu,  
  JCAP {\bf 1203} (2012) 016  [arXiv:1105.6226 [astro-ph.CO]].

\bibitem{Liao:2012ws}
  K.~Liao and Z.~H.~Zhu,  
  Phys.\ Lett.\ B {\bf 714} (2012) 1  [arXiv:1207.2552 [astro-ph.CO]].

\bibitem{Wu:2014wra}
  J.~Wu, Z.~Li, P.~Wu and H.~Yu,  
  Sci.\ China Phys.\ Mech.\ Astron.\  {\bf 57} (2014) 988.

\bibitem{Chen:2013vea}
  Y.~Chen, C.~Q.~Geng, S.~Cao, Y.~M.~Huang and Z.~H.~Zhu,  
  JCAP {\bf 1502} (2015) 02,  010  [arXiv:1312.1443 [astro-ph.CO]].

\bibitem{Cui:2015oda}
  J.~Cui, Y.~Xu, J.~Zhang and X.~Zhang,  
  Sci.\ China Phys.\ Mech.\ Astron.\  {\bf 58}, 110402 (2015)
  [arXiv:1511.06956 [astro-ph.CO]].

\bibitem{Ruff:2010rv}
  A.~J.~Ruff, R.~Gavazzi, P.~J.~Marshall, T.~Treu, M.~W.~Auger and F.~Brault,  
  Astrophys.\ J.\  {\bf 727}, 96 (2011) 
  [arXiv:1008.3167 [astro-ph.CO]].

\bibitem{Cao:2015qja}
  S.~Cao, M.~Biesiada, R.~Gavazzi, A.~Pi¨®rkowska and Z.~H.~Zhu,  
  Astrophys.\ J.\  {\bf 806}, 185 (2015) 
  [arXiv:1509.07649 [astro-ph.CO]].


\bibitem{Li:2015sla}
  X.~L.~Li, S.~Cao, X.~G.~Zheng, S.~Li and M.~Biesiada,  
  Res.\ Astron.\ Astrophys.\  {\bf 16}, no. 5, 084 (2016)  
  [arXiv:1510.03494 [astro-ph.CO]].

\bibitem{Guo:2015gpa}
  R.~Y.~Guo and X.~Zhang,  
  Eur.\ Phys.\ J.\ C {\bf 76}, no. 3, 163 (2016)  
  [arXiv:1512.07703 [astro-ph.CO]].

\bibitem{Li:2004rb}
  M.~Li,  
  Phys.\ Lett.\ B {\bf 603}, 1 (2004)  [hep-th/0403127].

  \bibitem{Gao:2007ep}
  C.~Gao, F. Q. Wu, X.~Chen and Y.~-G.~Shen,  ``A Holographic Dark Energy Model from Ricci Scalar Curvature,''
  Phys.\ Rev.\ D {\bf 79}, 043511 (2009)  [arXiv:0712.1394 [astro-ph]].

\bibitem{Bolton:2008xf}
  A.~S.~Bolton, S.~Burles, L.~V.~E.~Koopmans, T.~Treu, R.~Gavazzi, L.~A.~Moustakas, R.~Wayth and D.~J.~Schlegel,  
  Astrophys.\ J.\  {\bf 682}, 964 (2008)  
  [arXiv:0805.1931 [astro-ph]].

\bibitem{Auger:2009hj}
  M.~W.~Auger, T.~Treu, A.~S.~Bolton, R.~Gavazzi, L.~V.~E.~Koopmans, P.~J.~Marshall, K.~Bundy and L.~A.~Moustakas,  
  Astrophys.\ J.\  {\bf 705}, 1099 (2009)  
  [arXiv:0911.2471 [astro-ph.CO]].

\bibitem{Bolton:2011bj}
  J.~R.~Brownstein {\it et al.},  
  Astrophys.\ J.\  {\bf 744}, 41 (2012)  
  [arXiv:1112.3683 [astro-ph.CO]].

\bibitem{Treu:2002ee}
  T.~Treu and L.~Koopmans,  
  Astrophys.\ J.\  {\bf 575}, 87 (2002)  
  [astro-ph/0202342].

\bibitem{Koopmans:2002qh}
  L.~V.~E.~Koopmans and T.~Treu,  
  Astrophys.\ J.\  {\bf 583}, 606 (2003)  
  [astro-ph/0205281].

\bibitem{Treu:2004wt}
  T.~Treu and L.~V.~E.~Koopmans,  
  Astrophys.\ J.\  {\bf 611}, 739 (2004)  
  [astro-ph/0401373].

\bibitem{Sonnenfeld:2013cha}
  A.~Sonnenfeld, R.~Gavazzi, S.~H.~Suyu, T.~Treu and P.~J.~Marshall,  
  Astrophys.\ J.\  {\bf 777}, 97 (2013)  
  [arXiv:1307.4764 [astro-ph.CO]].

\bibitem{Sonnenfeld:2013xga}
  A.~Sonnenfeld, T.~Treu, R.~Gavazzi, S.~H.~Suyu, P.~J.~Marshall, M.~W.~Auger and C.~Nipoti,  
  Astrophys.\ J.\  {\bf 777}, 98 (2013)  
  [arXiv:1307.4759 [astro-ph.CO]].

\bibitem{Betoule:2014frx}
  M.~Betoule {\it et al.} [SDSS Collaboration],  
  Astron.\ Astrophys.\  {\bf 568}, A22 (2014)  
  [arXiv:1401.4064 [astro-ph.CO]].

\bibitem{Ade:2015rim}
  P.~A.~R.~Ade {\it et al.} [Planck Collaboration],  
  Astron.\ Astrophys.\  {\bf 594}, A14 (2016)  
  [arXiv:1502.01590 [astro-ph.CO]].

\bibitem{Hu:1995en}
  W.~Hu and N.~Sugiyama,  
  Astrophys.\ J.\  {\bf 471} (1996) 542  [astro-ph/9510117].

\bibitem{Eisenstein:1997ik}
  D.~J.~Eisenstein and W.~Hu,  
  Astrophys.\ J.\  {\bf 496} (1998) 605  [astro-ph/9709112].

\bibitem{Beutler:2011hx}
  F.~Beutler {\it et al.},  
  Mon.\ Not.\ Roy.\ Astron.\ Soc.\  {\bf 416}, 3017 (2011)  
  [arXiv:1106.3366 [astro-ph.CO]].

\bibitem{Ross:2014qpa}
  A.~J.~Ross, L.~Samushia, C.~Howlett, W.~J.~Percival, A.~Burden and M.~Manera,  
  Mon.\ Not.\ Roy.\ Astron.\ Soc.\  {\bf 449}, no. 1, 835 (2015)  
  [arXiv:1409.3242 [astro-ph.CO]].

\bibitem{Anderson:2013zyy}
  L.~Anderson {\it et al.} [BOSS Collaboration],  
  Mon.\ Not.\ Roy.\ Astron.\ Soc.\  {\bf 441}, no. 1, 24 (2014)  
  [arXiv:1312.4877 [astro-ph.CO]].



\bibitem{Xu:2016grp}
  Y.~Y.~Xu and X.~Zhang,
  Eur.\ Phys.\ J.\ C {\bf 76}, no. 11, 588 (2016)
  [arXiv:1607.06262 [astro-ph.CO]].

\bibitem{Cheng:2014kja} 
  C.~Cheng and Q.~G.~Huang,
  Sci.\ China Phys.\ Mech.\ Astron.\  {\bf 58}, no. 9, 599801 (2015)
  [arXiv:1409.6119 [astro-ph.CO]].

\bibitem{Hu:2015bpa} 
  S.~Wang, M.~Li and Y.~Hu,
  Sci.\ China Phys.\ Mech.\ Astron.\  {\bf 60}, no. 4, 040411 (2017)
  [arXiv:1506.08274 [astro-ph.CO]].



\bibitem{Blake:2012pj}
  C.~Blake {\it et al.},  
  Mon.\ Not.\ Roy.\ Astron.\ Soc.\  {\bf 425}, 405 (2012)  
  [arXiv:1204.3674 [astro-ph.CO]].

\bibitem{Chuang:2012qt}
  C.~H.~Chuang and Y.~Wang,  
  Mon.\ Not.\ Roy.\ Astron.\ Soc.\  {\bf 435}, 255 (2013)  
  [arXiv:1209.0210 [astro-ph.CO]].

\bibitem{Delubac:2014aqe}
  T.~Delubac {\it et al.} [BOSS Collaboration],  
  Astron.\ Astrophys.\  {\bf 574}, A59 (2015)  
  [arXiv:1404.1801 [astro-ph.CO]].

\bibitem{Stern:2009ep}
  D.~Stern, R.~Jimenez, L.~Verde, M.~Kamionkowski and S.~A.~Stanford,  
  JCAP {\bf 1002}, 008 (2010)  
  [arXiv:0907.3149 [astro-ph.CO]].

\bibitem{Moresco:2012jh}
  M.~Moresco {\it et al.},  
  JCAP {\bf 1208}, 006 (2012)  
  [arXiv:1201.3609 [astro-ph.CO]].

\bibitem{Zhang:2012mp}
  C.~Zhang, H.~Zhang, S.~Yuan, T.~J.~Zhang and Y.~C.~Sun,  
  Res.\ Astron.\ Astrophys.\  {\bf 14}, no. 10, 1221 (2014)  
  [arXiv:1207.4541 [astro-ph.CO]].

\bibitem{Moresco:2015cya}
  M.~Moresco,  
  Mon.\ Not.\ Roy.\ Astron.\ Soc.\  {\bf 450}, no. 1, L16 (2015)  
  [arXiv:1503.01116 [astro-ph.CO]].

\bibitem{Ade:2015xua}
  P.~A.~R.~Ade {\it et al.} [Planck Collaboration],  
  Astron.\ Astrophys.\  {\bf 594}, A13 (2016)  
  [arXiv:1502.01589 [astro-ph.CO]].

\bibitem{Zhang:2005yz} 
  X.~Zhang,
  Int.\ J.\ Mod.\ Phys.\ D {\bf 14}, 1597 (2005)
  [astro-ph/0504586].

\bibitem{Zhang:2005hs} 
  X.~Zhang and F.~Q.~Wu,
  Phys.\ Rev.\ D {\bf 72}, 043524 (2005)
  [astro-ph/0506310].

\bibitem{Zhang:2007sh} 
  X.~Zhang and F.~Q.~Wu,
  Phys.\ Rev.\ D {\bf 76}, 023502 (2007)
  [astro-ph/0701405].

\bibitem{Zhang:2006av} 
  X.~Zhang,
  Phys.\ Lett.\ B {\bf 648}, 1 (2007)
  [astro-ph/0604484].


\bibitem{Zhang:2006qu} 
  X.~Zhang,
  Phys.\ Rev.\ D {\bf 74}, 103505 (2006)
  [astro-ph/0609699].



\bibitem{Zhang:2009xj} 
  X.~Zhang,
  Phys.\ Lett.\ B {\bf 683}, 81 (2010)
  [arXiv:0909.4940 [gr-qc]].

\bibitem{Wang:2013zca} 
  S.~Wang, J.~J.~Geng, Y.~L.~Hu and X.~Zhang,
  Sci.\ China Phys.\ Mech.\ Astron.\  {\bf 58}, no. 1, 019801 (2015)
  [arXiv:1312.0184 [astro-ph.CO]].

\bibitem{Zhou:2016rtz} 
  L.~Zhou and S.~Wang,
  Sci.\ China Phys.\ Mech.\ Astron.\  {\bf 59}, no. 7, 670411 (2016)
  [arXiv:1602.02213 [astro-ph.CO]].

\bibitem{Zhang:2015uhk} 
  X.~Zhang,
  Phys.\ Rev.\ D {\bf 93}, no. 8, 083011 (2016)
  [arXiv:1511.02651 [astro-ph.CO]].

\bibitem{Wang:2016tsz} 
  S.~Wang, Y.~F.~Wang, D.~M.~Xia and X.~Zhang,
  Phys.\ Rev.\ D {\bf 94}, no. 8, 083519 (2016)
  [arXiv:1608.00672 [astro-ph.CO]].

\bibitem{Zhang:2017rbg} 
  X.~Zhang,
  Sci.\ China Phys.\ Mech.\ Astron.\  {\bf 60}, no. 6, 060431 (2017)
  [arXiv:1703.00651 [astro-ph.CO]].

\bibitem{He:2016rvp} 
  D.~Z.~He, J.~F.~Zhang and X.~Zhang,
  Sci.\ China Phys.\ Mech.\ Astron.\  {\bf 60}, no. 3, 039511 (2017)
  [arXiv:1607.05643 [astro-ph.CO]].



\bibitem{Zhang:2009un} 
  X.~Zhang,
  Phys.\ Rev.\ D {\bf 79}, 103509 (2009)
  [arXiv:0901.2262 [astro-ph.CO]].

\bibitem{Feng:2009hr} 
  C.~J.~Feng and X.~Zhang,
  Phys.\ Lett.\ B {\bf 680}, 399 (2009)
  [arXiv:0904.0045 [gr-qc]].

\bibitem{Li:2009bn} 
  M.~Li, X.~D.~Li, S.~Wang and X.~Zhang,
  JCAP {\bf 0906}, 036 (2009)
  [arXiv:0904.0928 [astro-ph.CO]].


\bibitem{Zhang:2010im} 
  J.~Zhang, L.~Zhang and X.~Zhang,
  Phys.\ Lett.\ B {\bf 691}, 11 (2010)
  [arXiv:1006.1738 [astro-ph.CO]].

\bibitem{Fu:2011ab} 
  T.~F.~Fu, J.~F.~Zhang, J.~Q.~Chen and X.~Zhang,
  Eur.\ Phys.\ J.\ C {\bf 72}, 1932 (2012)
  [arXiv:1112.2350 [astro-ph.CO]].

\bibitem{Zhang:2014sqa} 
  J.~F.~Zhang, J.~L.~Cui and X.~Zhang,
  Eur.\ Phys.\ J.\ C {\bf 74}, no. 10, 3100 (2014)
  [arXiv:1409.6562 [astro-ph.CO]].













  \end{thebibliography}
\end{document}